\documentclass[
 reprint,
superscriptaddress,
 amsmath,amssymb
]{revtex4-2}
\usepackage{tabularx}
\newcolumntype{C}[1]{>{\centering\let\newline\\\arraybackslash\hspace{0pt}}m{#1}}
\usepackage{bbold}
\usepackage{xcolor}
\usepackage{graphicx}
\usepackage{dcolumn}
\usepackage{bm}

\usepackage{float}
\usepackage{lipsum}

\begin{document}

\preprint{APS/123-QED}

\title{A Coarse-grained Model for Aqueous Two-phase Systems: Application to Ferrofluids}

\author{Alberto Scacchi}\email{alberto.scacchi@aalto.fi}
\affiliation{Department of Applied Physics, Aalto University, P.O. Box 11000, FI-00076 Aalto, Finland}
\affiliation{Academy of Finland Center of Excellence in Life-Inspired Hybrid Materials (LIBER), Aalto University, P.O. Box 16100, FI-00076 Aalto, Finland}

\author{Carlo Rigoni}
\affiliation{Department of Applied Physics, Aalto University, P.O. Box 11000, FI-00076 Aalto, Finland}
\affiliation{Academy of Finland Center of Excellence in Life-Inspired Hybrid Materials (LIBER), Aalto University, P.O. Box 16100, FI-00076 Aalto, Finland}

\author{Mikko Haataja}
\affiliation{Department of Mechanical and Aerospace Engineering, and Princeton Materials Institute (PMI),
Princeton University, Princeton, New Jersey 08544, United States}

\author{Jaakko V. I. Timonen}
\affiliation{Department of Applied Physics, Aalto University, P.O. Box 11000, FI-00076 Aalto, Finland}
\affiliation{Academy of Finland Center of Excellence in Life-Inspired Hybrid Materials (LIBER), Aalto University, P.O. Box 16100, FI-00076 Aalto, Finland}

\author{Maria Sammalkorpi}
\affiliation{Academy of Finland Center of Excellence in Life-Inspired Hybrid Materials (LIBER), Aalto University, P.O. Box 16100, FI-00076 Aalto, Finland}
\affiliation{Department of Chemistry and Materials Science, Aalto University, P.O. Box 16100, FI-00076 Aalto, Finland}
\affiliation{Department of Bioproducts and Biosystems, Aalto University, P.O. Box 16100, FI-00076 Aalto, Finland}

\date{\today}

\begin{abstract}
Aqueous two-phase systems (ATPSs), that is, phase-separating solutions of water soluble but mutually immiscible molecular species, offer fascinating prospects for selective partitioning, purification, and extraction. Here, we formulate a general Brownian dynamics based coarse-grained simulation model for a polymeric ATPS comprising two water soluble but mutually immiscible polymer species. A third solute species, representing, e.g., nanoparticles (NPs), additional macromolecular species, or impurities can readily be incorporated into the model.
We demonstrate that the model captures satisfactorily the phase separation, partitioning, and interfacial properties of a model ATPS composed of a polymer mixture of dextran and polyethylene glycol (PEG) in which magnetic NPs selectively partition into one of the two polymeric phases. The NP partitioning is characterized both via the computational model and experimentally under different conditions. The simulation model captures the trends observed in the experiments and quantitatively links the partitioning behavior to the component species interactions. Finally, the response of the simulation model to external magnetic field, with the magnetic NPs as the additional partitioned component, shows that the ATPS interface fluctuations can be controlled by the magnetic field at length scales much smaller than those probed experimentally to date.
\end{abstract}

\maketitle

\section{Introduction}

Phase separation of macromolecular solutions resulting in coexisting aqueous phases underlies the formation of membraneless compartments in biological cells~\cite{brangwynne2015polymer,hyman2014liquid, alberti2019liquid, alberti2019considerations}, gives rise to advanced functionalities in polymeric materials~\cite{wang2019progress}, and enables a broad range of technological applications related to biomolecules and bio species, including separation and purification of nucleic acids and proteins (enzymes and antibodies) as well as extraction of viruses~\cite{raja2011aqueous,asenjo2011aqueous,atefi2015interfacial}. Furthermore, aqueous phase-separating solutions also hold the promise for low environmental impact extraction techniques~\cite{assis2021application} and find broad usage in biomedicine and biopharmaceuticals~\cite{chao2020emerging,rosa2010aqueous}, as well as in biotechnology~\cite{raja2011aqueous,walter1986partitioning, teixeira2018emerging}. Phase-separating systems can also guide cellular micropatterning and encapsulation to mimic biological cells~\cite{pereira2020aqueous}. Also, several pathological conditions, including those associated with neurological diseases~\cite{Zbinden2020} or cancer spreading~\cite{jiang2020protein}, may be affected by the phase-separation processes.

At a general level, ATPSs form when incompatible polymers and/or salts are mixed with water such that phase separation in the solution  arises~\cite{hatti2001aqueous,iqbal2016aqueous} with each phase enriched in one of the two components.
In a prior work, we examined a polymer-based ATPS experimentally, in the context of a ferrofluid system~\cite{rigoni2022ferrofluidic}.
Ferrofluids, comprising of suspensions of nanoscale particles in fluids, allow for reversible pattern formation response under external fields~\cite{zhang2019flexible}. Indeed, their magnetically~\cite{maiorov1983magnetic,torres2014recent, lee2017magnetoresponsive, zou2018composite} and electrically~\cite{cherian2021electroferrofluids} driven control underlie a variety of advanced technological applications~\cite{Kole2021, xia2010ferrofluids,zhang2019flexible}. 
The magnetically responsive NPs in polymer solutions also have broad use in biomedical technologies, e.g., with various drug delivery~\cite{ulbrich2016targeted}, including targeted thermochemotherapy~\cite{hayashi2016smart}, approaches. Likewise, the enhanced thermal conductivity also allows for energy storage and generation applications of NP based ferrofluids~\cite{Kole2021,afifah2016magnetoviscous}.
 
Since phase separation of polymer mixtures commonly involves length and time scales larger than micrometers and microseconds, respectively, theoretical and computational modelling approaches require coarse-graining of the molecular level details. Typical coarse-grained particle-based modelling approaches for polymer systems include, e.g., dissipative particle dynamics~\cite{espanol2017perspective} and Brownian or Langevin dynamics simulations~\cite{huber2019brownian}. Free energy-based field-theoretic approaches include, e.g., self-consistent field theory~\cite{muller2013, muller2020}, classical density functional theory~\cite{lutsko2010recent,te2020classical}, and power functional theory~\cite{schmidt2022power}. However, developing predictive field-based approaches is challenging for complex systems. Additionally, field-based approaches typically provide only expectation values and distributions, rather than localized, particle or molecule level information. 

In this work, we combine theoretical/simulational and experimental methods to study a polymer based ATPS. Specifically, we examine the system phase separation response, generic partitioning within, and the response to external magnetic field of magnetic NPs dispersed in the system. To this end, we construct a Brownian dynamics based simulation model to describe the ATPS and systematically connect and compare the model to an ATPS introduced and experimentally characterized by us in Ref.~\citenum{rigoni2022ferrofluidic}. Notably, the coarse-grained, particle-based approach allows accessing time and length scales of the ATPS partitioning and dynamics response. 
We demonstrate the model and its ability to predict the response of ATPSs by comparing the phase separation and partitioning response predicted by the simulations model and measured experimentally for actual ATPSs with varying characteristics. Throughout, the model construction, verification steps, and the parametrization are presented in detail such that our approach can be readily adapted for other ATPSs. 

\section{Materials and methods}

\subsection*{Theory}\label{sec::brownian}
We construct a Brownian dynamics simulations model for aqueous solution of two polymers and magnetic NPs. Notably, as the NPs are magnetic, also magnetic dipole interactions need to be considered. Brownian dynamics is an implicit-solvent particle-based modelling method in which the solutes constitute the particles in the simulations model, and are considered to be large in comparison to the solvent molecules. The effects rising from the solvent on the solutes are captured via an effective random force term. The approach leads to a computationally very efficient effective description of the solvent influence, allowing extensive length and time scales in modelling compared to e.g., atomistic, or even coarse-grained detailed solvent modelling approaches. 

Each particle follows a time trajectory resulting from the equations of motion of the position vectors $\textbf{r}_i$:
\begin{equation}
    \textbf{r}_i(t+dt)=\textbf{r}_i(t)-\frac{\nabla\Phi(\textbf{r}_i,t)}{\Gamma_{\rm t}} dt +\delta \textbf{r}_i.
\label{eq:bd}
\end{equation}
Here, the total interaction potential $\Phi$ includes all the one- and two-body interaction contributions to the solute interactions, and $\Gamma_{\rm t}$ is the translational friction constant (here set to be the same for each particle). The time $t$ is propagated in discrete steps $dt$, the integration time step. 
The three components of the stochastic position change $\delta \textbf{r}_i$ are sampled from a Gaussian distribution with standard deviation $\sqrt{2D_{\rm t} dt}$, where  $D_{\rm t}$ the translational diffusion constant, which follows the Einstein relation $D_{\rm t}=k_{\rm B}T/\Gamma_{\rm t}$. Here $k_{\rm B}$ is the Boltzmann constant and $T$ the temperature. $\beta^{-1}=k_{\rm B}T$ defines the energy scale of the simulations, and is set to 1 in this work. Without loss of generality, we also fix $D_{\rm t}=1$; this choice sets the time scale of the simulations, and can be used to convert modelling time scales to real-time equivalents.

The magnetic dipole moments of the NPs, $\boldsymbol{\mu}_i$, evolve in time as
\begin{equation}
    \boldsymbol{\mu}_i(t+dt)=\frac{\boldsymbol{\mu}_i(t)+\boldsymbol{\omega}_i\times \boldsymbol{\mu}_i dt}{\vert \boldsymbol{\mu}_i(t)+\boldsymbol{\omega}_i\times \boldsymbol{\mu}_i\vert},
\label{eq:dipolebd}
\end{equation}
where 
\begin{equation}
    \boldsymbol{\omega}_i=\frac{\textbf{T}_i}{\Gamma_r}+\delta \boldsymbol{\omega}_i.
\end{equation}
The three components of the stochastic rotation, $\delta \boldsymbol{\omega}_i$, are sampled from a Gaussian distribution with standard deviation $\sqrt{2D_{\rm r}}$, where $D_{\rm r}$ is the rotational diffusion constant. Also $D_{\rm r}$ and the corresponding friction coefficient $\Gamma_{\rm r}$ connect by the Einstein relation $D_{\rm r}=k_{\rm B}T/\Gamma_{\rm r}$. $\Gamma_{\rm t}$ and $\Gamma_{\rm r}$ at infinite dilution with stick boundary conditions follow the relation $\Gamma_{\rm t}=3\Gamma_{\rm r}/d^2$, where $d$ is the diameter of the particle. 
Finally, $\textbf{T}_i$ describes all torques acting on particle $i$, which arise from dipole-dipole interactions or from interaction with external fields.

Brownian dynamics-type approaches are valid only if the modelled solute particles are much larger than the solvent molecules, and if the velocity-velocity correlation decays on a time scale much smaller than the simulation time scale, often referred to as viscous limit. Notably, both these conditions are satisfied by the ATPS characterized in Ref.~\citenum{rigoni2022ferrofluidic}, see Appendix~\ref{app::viscous_limit}. If the latter condition is not fulfilled, Langevin dynamics~\cite{langevin1908} is a more appropriate choice than Brownian dynamics.

\subsection*{Simulations and software}

The Large-scale Atomic/Molecular Massively Parallel Simulator (LAMMPS)~\cite{thompson2022lammps} {\it 23 Jun 2022} stable version with an optimised interaction scheme~\cite{in2008accurate, shire2021simulations} is used for the simulations. 
All initial configurations of this work are prepared using Packmol~\cite{martinez2009packmol} and Moltemplate~\cite{jewett2021moltemplate}. The visualizations are produced using either VMD~\cite{vmd} or Ovito~\cite{stukowski2009visualization}. 

\subsection*{Materials and experimental characterization}
The mixtures of NPs, PEG, and dextran in water solution have been prepared using the following procedure in four steps.
First, two separate concentrated solutions of PEG (polyethylene glycol 35000, $\geq$ 99.9 \%, 81310, Sigma-Aldrich) and dextran (Dextran T500, $\sim$ 95 \%, 40030,  Pharmacosmos) were obtained by mixing 75.0 g of polymers with 150.0 g of deionized water (Ultrapure type 1 water obtained with Millipore Direct-Q 3UV). The final densities of the PEG and dextran solutions, measured as in Ref.~\citenum{rigoni2022ferrofluidic}, are $1.052\pm 0.003$~g/ml and $1.078 \pm 0.001$~g/ml, respectively.
Second, a main ATPS dispersion was prepared by mixing 2635.2 mg of PEG concentrated solution, 5832.7 mg of dextran concentrated solution, and 20794.7 mg of deionized water.
Third, in order to obtain approximately similar final concentrations of iron oxide NPs in all samples, the stock NPs dispersions were diluted using a micropipette (Eppendorf Research Plus) in the following way: 15 $\mu$l of stock solution of citrated coated iron oxide NPs (see synthesis details in Ref.~\citenum{rigoni2022ferrofluidic}) + 985 $\mu$l of deionized water, 238 $\mu$l of phospholipid coated iron oxide NPs (45-111-701, Micromod) + 762 $\mu$l of deionized water, 6 $\mu$l of pegylated iron oxide NPs (PBG 300, Ferrotec) + 994 $\mu$l of deionized water.
Finally, the NPs-PEG-dextran dispersions were prepared using the following methodology: the main ATPS dispersion was shaken in order to obtain an emulsion from which 650 $\mu$l were collected using a micropipette, and poured in the final container. Similarly the diluted NPs dispersions were shaken, and 100 $\mu$l were then collected with a micropipette and poured in the final container. An additional control ATPS was obtained by adding 100 $\mu$l of deionized water instead of the NPs dispersion. All dispersions were prepared in duplicates in glass vials (Supelco 29403-U) for the photographs and in poly(methyl methacrylate) (PMMA) cuvettes (VWR Cuvettes PMMA semi-micro 634-0678) for the spectrophotometry experiments. All dispersions were then mixed and let rest for 48 hours before performing the experiments and taking the pictures, which were collected using a Nikon digital camera D5500. Optical absorption of the samples was measured using a spectrophotometer (Thermo Scientific GENESYS 30). The PMMA cuvettes filled with the samples were placed at different height inside the spectrophotometer to probe the light absorption as a function of the wavelength in the dilute and in the dense phases.

\section{Results and discussion}

\subsection*{ATPS computational model construction and partitioning results}

To enable further development work, the simulation model construction is presented modularly with four levels of complexity. First, (i) a binary polymer mixture that exhibits phase separation is considered. Next, we include (ii) a third (magnetically responsive) component with asymmetric interactions with respect to the two polymer species, (iii) consider the effect of gravity, and (iv) investigate the effects of an external magnetic field.

\subsubsection*{Polymer mixture}\label{sec::pol_mix}
To construct the ATPS model, we first consider two different polymer species in an aqueous solution. The Brownian dynamics equations of motion, Eq.~(\ref{eq:bd}), require determining the pair interaction potential for the polymer species.
Gaussian-like potentials between centers of mass have been found to describe the effective interaction between both linear polymers and dendrimers in good solvents~\cite{louis2000can, bolhuis2001accurate, likos2001effective, gotze2004tunable, likos2006soft, gotze2006structure}. Supported by this, we choose to describe the interaction of the binary mixture of polymers by 
\begin{equation}
\label{eq::gaussian}
\phi_{ij}(r)=\varepsilon_{ij}e^{-r^2/R_{ij}^2},\quad i,j=1,2,
\end{equation}
where $r$ is the center of mass-center of mass separation, $R_{ij}$ can be connected with the radius of gyration, and $\varepsilon_{ij}$ denote interaction strengths. The sub-indices $i$ and $j$ refer to the species type.
In particular, Louis {\it et al.}~\cite{louis2000can} have shown that such Gaussian-like potential with a pre-factor $\varepsilon \approx 2k_{\rm B}T$ describe the interaction between linear polymers in a good solvent very well over a wide range of concentrations. 

The comparison system of Ref.~\cite{rigoni2022ferrofluidic} contains, in an aqueous solution, PEG (``P$_{1}$''), and dextran (``P$_{2}$''). PEG is a linear polymer for which water is a good solvent. Thus, setting $\varepsilon_{11}=2k_{\rm B}T$ is well motivated. On the other hand, dextran is a branched polymer for which water is rather a $\theta$-solvent, resulting in dextran coiling~\cite{antoniou2012solution}. Nevertheless, we set $\varepsilon_{22}=2k_{\rm B}T$ also for dextran for simplicity, as no generic functional form of the effective interaction between branched polymers in $\theta$-solvent exists to the best of our knowledge. 
Additionally, we set $R_{11}=R_{12}=R_{22}=R$, even though in Ref.~\citenum{rigoni2022ferrofluidic} the radius of gyration of dextran is about twice as large as PEG. The remaining variable is $\varepsilon_{12}$, which controls the miscibility of the polymers, and is unknown for the experimental reference system of Ref.~\cite{rigoni2022ferrofluidic}. However, as the reference system is an ATPS, phase separation of the two polymer species is expected. 

The condition for phase separation of two polymeric species interacting following Eq.~(\ref{eq::gaussian}) at fixed volume is~\cite{louis2000mean}
\begin{equation}
    \chi = \beta\pi^{3/2}\left[2\varepsilon_{12}R_{12}^3-\left(\varepsilon_{11}R_{11}^3+\varepsilon_{22}R_{22}^3\right)\right]>0.
    \label{eq:phasesep}
\end{equation}
The spinodal number density $\rho_{\rm s}(x)$ indicating the critical number density for phase separation is~\cite{louis2000mean}
\begin{equation}
    \rho_{\rm s}(x)=\left(x(1-x)\chi\right)^{-1}.
\end{equation}
\begin{figure*}[!htb]
    \centering
    \includegraphics[width=\linewidth]{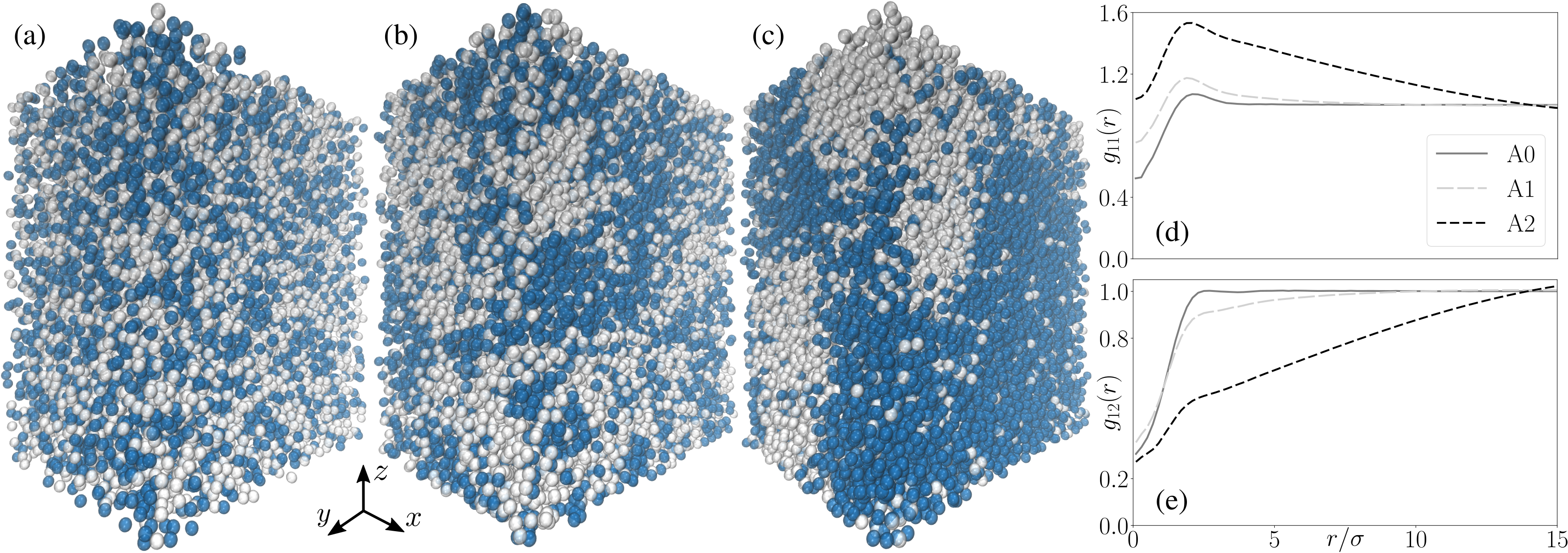}
    \caption{Assembly configurations rising in modelling polymer - polymer systems at different concentrations, and the corresponding homogeneous and isotropic radial distribution functions. Panels correspond to visualizations of the assembly structure in systems (a) A0, (b) A1, and (c) A2 after $200\times 10^6$ time steps. P$_1$ are grey and P$_2$ blue. Panels (d) and (e) show the radial distribution functions $g_{11}$ and $g_{12}$, see Eq.~(\ref{eq::g}).}
    \label{figure::1}
\end{figure*}
Here $x=N_2/N_{\rm tot}$, where $N_2$ is the number of P$_{2}$ polymers and $N_{\rm tot}=N_1+N_2$ the total number of polymers in the system. The choices made above for the interactions of P$_1$ and P$_2$ result in the phase separation criterion ($\chi > 0$ in Eq.~(\ref{eq:phasesep})) for the modelled system becoming $\varepsilon_{12}>2k_{\rm B}T$. To satisfy this, we choose  $\varepsilon_{12}=2.5k_{\rm B}T$ for the model. A thorough study on the mixing and phase separation response under different polymer interaction conditions, $\varepsilon_{ij}$ and $R_{ij}$, was presented in Ref.~\citenum{archer2001binary}. The choice of setting $R_{ij}=R\quad\forall i,j$ results in the most symmetric phase response (see Fig.~5 in Ref.~\cite{archer2001binary}).
\begin{table}[h!]
\centering
\begin{tabular}{ C{1.2cm} C{1.2cm} C{1.2cm} C{1.2cm} C{1.2cm} C{1.2cm}}
System & $N_{\rm tot}$ & $N_{\rm NP}$ & $\rho_{\rm tot}\sigma^3$ & $\rho_{\rm NP}\sigma^3$ & $\beta\varepsilon_{23}$\\  \hline\hline
A0 & 12000 & - & 0.222 & - & -\\ \hline
A1 & 20000 & - & 0.370 & - & -\\ \hline
A2 & 25000 & - & 0.463 & - & -\\ \hline
A3 & 30000 & - & 0.556 & - & -\\ \hline
B1 & 20000 & 5000 & 0.370 & 0.093 & 1\\ \hline
B2 & 20000 & 5000 & 0.370 & 0.093 & 3\\ \hline
B3 & 20000 & 5000 & 0.370 & 0.093 & 4\\ \hline
C1 & 25000 & 5000 & 0.463 & 0.093 & 1\\ \hline
C2 & 25000 & 5000 & 0.463 & 0.093 & 2\\ \hline
C3 & 25000 & 5000 & 0.463 & 0.093 & 3\\
\end{tabular}
\caption{Summary of the system nomenclature and the different examined conditions. The columns present for each system the total number of polymers $N_{\rm tot}$, number of nanoparticles $N_{\rm NP}$, total polymer number density $\rho_{\rm tot}\sigma^3$, total nanoparticles number density $\rho_{\rm NP}\sigma^3$, and attraction strength between P$_2$ and NPs $\beta\varepsilon_{23}$.} \label{table::1}
\end{table}

In the simulations, Eq.~(\ref{eq::gaussian}) is truncated and shifted to zero at $r=3.5\sigma$. Here, $\sigma=R/\sqrt{2}$ defines the length-scale of the simulations. We consider in the modelling a simulation volume $V= L_{x} \times L_{y} \times L_{z} = 30\sigma \times 40\sigma \times 45\sigma$. The critical number of polymers for phase separation to occur is $N_{\rm c}=\rho_{\rm s} V$. For $x=0.5$, i.e., $N_1=N_2$, in the simulation volume, this corresponds to polymer number count of 13715, or calculated as number density, to $\rho_{\rm s}\sigma^3\approx 0.254$. To elucidate the possible system responses with concentration changes, three different polymer concentrations are compared: system A0 corresponds to $N_{\rm tot}=12000<N_{\rm c}$, system A1 to $N_{\rm tot}=20000>N_{\rm c}$, and system A2 to $N_{\rm tot}=25000>N_{\rm c}$ polymers. The examined systems correspond to number densities $\rho_{\rm tot}\sigma^3\approx0.222$, 0.370, and 0.463, respectively. The different system parameters are summarized in Table~\ref{table::1}.

The experimental comparison system of Ref.~\citenum{rigoni2022ferrofluidic} contains in the characterized {\it Sample 0}, and in an equivalent volume, approximately $19850$ polymers (calculated assuming $\sigma=10$ nm, based on the magnitude of the average size of the three components in the solution in the experiments). Of these, $2630$ are dextran and $17220$ are PEG, based on their concentrations. This corresponds to a total number density $\rho_{\rm tot}^{\rm exp}\sigma^3\approx 0.368$, to which dextran contributes $0.049$ and PEG $0.319$. A detailed description of the calculation of these estimates is shown in Appendix~\ref{app::numbers}.

All simulations have as their initial configurations a random mixture of the polymer particles in the given particle number density. Periodic boundary conditions along the $xy$ plane, but reflective hard walls in the $z$ direction, are employed. The wall boundary along the $z$ direction is expected to slightly affect the bulk phase response, as well as the critical density $\rho_{\rm s}$. The wall boundary is used to allow setting up a more realistic system for the later stages of this work, where gravity-like conditions are also considered.

Figure~\ref{figure::1} shows the assembly configurations rising for the different polymer concentrations systems A0, A1, and A2, both as simulation visualizations and corresponding homogeneous and isotropic radial distribution functions (see, e.g., Sec.~2.5 of Ref.~\citenum{hansen2013theory})
\begin{equation}\label{eq::g}
g_{ij} (r) = \frac{\rho_{ij}(r)V'}{\int_{V'} dr 4 \pi r^{2} \rho_{ij}(r)}.
\end{equation}
Here we calculate $\rho_{ij}(r)$ by radially considering a shell of thickness $dr$ around particle $i$ at a distance $r$ from its center. The number of $j$ particles in the shell is counted and the result normalized by the shell volume. The denominator corresponds to the number of $j$ particles in the considered volume $V'$. Presented configurations in Fig.~\ref{figure::1} correspond to assembly structure at a simulation time of $200\times 10^6\> dt$, where $dt=10^{-5}\tau$. $\tau$ defines the simulation time scale by $\tau=\sigma^2/D_{\rm t}$. The analysis time period for calculating $g_{ij}(r)$ is the last 10$\%$ of the simulation time, and encompasses all particles in the system.   

As expected, for the chosen polymer interaction parameters and concentrations, phase separation rises clearly in A1 and A2, with the effect being more pronounced in A2. This is expected, since this configuration is deeper in the spinodal region, resulting in rather different coexisting densities than in A1. On the other hand, in A0, the polymer concentration is below the critical concentration. Thus, quite expectedly, the system does not show any clear sign of phase separation. 

The visual results are confirmed by the radial distribution functions data. The slow decay of $g_{11}(r)$ for both A1 and A2 in panel (d) of Fig.~\ref{figure::1} is an indication of phase separation in the systems. The same result is expected for $g_{22}(r)$, due to symmetry in the model setup. For A0, the slow decay of $g_{11}(r)$ is absent, as expected. Consistent with the intraspecies homogeneous radial distribution function, also the interspecies radial distribution function $g_{12}$, panel (e), shows phase-separation for A1 and A2 systems. For the interspecies radial distribution function, phase separation leads to a depletion for short-range and intermediate values of $r$. 

The observed response is at qualitative level consistent with the phase separations differences reported in the experiments of Ref.~\citenum{rigoni2022ferrofluidic}: Rigoni {\it et al.} present in Fig. 2b of Ref.~\cite{rigoni2022ferrofluidic} a comparison of phase separation response for PEG-dextran-iron oxide NPs systems with various levels of dilution. Out of the examined systems, four exhibit phase separation, and one remains in a mixed state, with concentration trend following the current work.
\begin{figure*}[!htb]
    \centering
    \includegraphics[width=\linewidth]{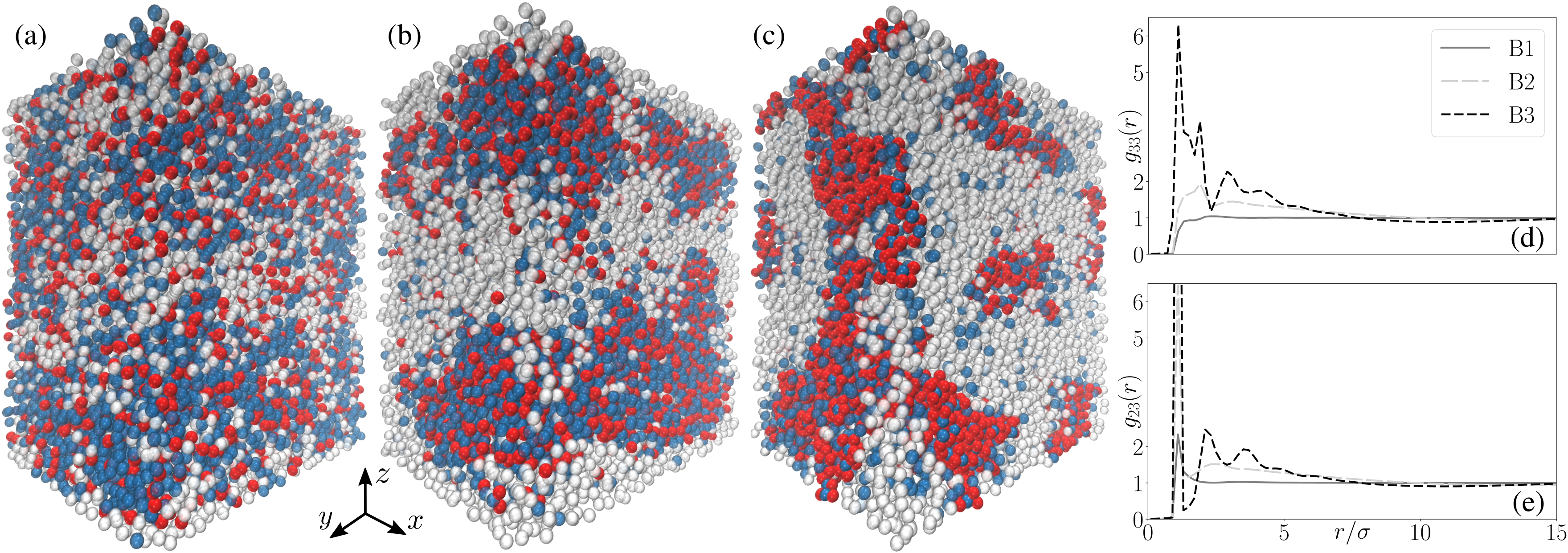}
    \caption{Assembly configurations rising in modelling polymer - polymer - NP systems with different $\beta\varepsilon_{\rm 23}$ values, and the corresponding homogeneous and isotropic radial distribution functions for the third species and its affinity polymer. Panels correspond to visualizations of the assembly structure in system (a) B1 ($\beta\varepsilon_{\rm 23}=1$), (b) B2 ($\beta\varepsilon_{\rm 23}=2$), and (c) B3 ($\beta\varepsilon_{\rm 23}=4$) after $50\times 10^6$ time steps. The value of $\beta\varepsilon_{\rm 23}$ indicates attraction strengths between P$_2$ and the NPs. P$_1$ are grey, P$_2$ blue, and NPs red. Panels (d) and (e) show the radial distribution functions $g_{33}$ and $g_{23}$, see Eq.~(\ref{eq::g}).}
    \label{figure::2}
\end{figure*}
\subsubsection*{Polymer mixture and nanoparticles}\label{sec::mix_and_np}
Let us now introduce a third solute component into the phase separating polymer mixture. Generally, this component can be e.g., biomacromolecules, such as nucleic
acids or proteins, including enzymes and antibodies, but also virus particles and pathogenic species, as well as colloidal particles or additional polymer species -- basically anything comparable in size to P$_1$ and P$_2$, and dispersed in aqueous solutions. Notably, the formulation is not restricted to a single additional component, yet for simplicity we focus here only on one additional species. Furthermore, to demonstrate how diverse types of interactions are introduced to the model, we choose magnetic NPs as the third component. Additionally, this allows further connection with the experimental ferrofluid system of Ref.~\citenum{rigoni2022ferrofluidic}. The interactions referring to the NP species is indicated by subindex 3, following the polymer components being 1 and 2. The pairwise interaction between NPs is modelled by $\phi^{\rm tot}_{33}=\phi_{\rm E}+\phi_{\rm WCA}+\phi_{\rm D}$. The first term, describing screened electrostatic interaction, is defined as
\begin{equation}\label{eq::electro}
    \phi_{\rm E}(r)=\frac{A}{\kappa}e^{-\kappa(r-\sigma_{3})},
\end{equation}
where $A>0$ controls the strength of the repulsive interactions, $\kappa$ is the inverse screening length, setting the range of the interactions, and $\sigma_{3}$ defines the diameter of the NPs. For simplicity, we choose $\sigma_{3}=\sigma$, and set $\beta\sigma A=25$ and $\sigma\kappa=5$. These choices provide a stable dispersion response of the NPs in the aqueous solution (colloidal behaviour). $\phi_{\rm E}$ is truncated and shifted to zero at $r=5\sigma$. The size of the NPs is defined by a standard Weeks-Chandler-Andersen (WCA) potential~\cite{weeks1971role}
\begin{equation}\label{WCA}
\phi_{\rm WCA}(r)=4\varepsilon_{3}\left[\left(\frac{\sigma}{r}\right)^{12} - \left( \frac{\sigma}{r}\right)^6\right]+\varepsilon_{3},\quad r < 2^{\frac{1}{6}} \sigma,  
\end{equation}
where $\varepsilon_{3}=1$. The ``shifted-force'' dipole-dipole interaction is given by~\cite{tildesley1987computer}
\begin{equation}
\begin{split}
\phi_{\rm D}(\textbf{r},\boldsymbol{\mu}_i, \boldsymbol{\mu}_j) &=  \left[1-4\left(\frac{r}{r_{\rm c}}\right)^{\!3}+
3\left(\frac{r}{r_{\rm c}}\right)^{\!4}\right]\\&\times \left[\frac{1}{r^3}(\boldsymbol{\mu}_i \cdot \boldsymbol{\mu}_j) - \frac{3}{r^5}
(\boldsymbol{\mu}_i \cdot \textbf{r}) (\boldsymbol{\mu}_j \cdot \textbf{r})\right],
\end{split}
\end{equation}
where $r=\vert\textbf{r}\vert $ is the distance between the centers of two NPs, and $\boldsymbol{\mu}_i$ and $\boldsymbol{\mu}_j$ their dipole moment vectors. $\phi_{\rm D}$, as the computationally most demanding interaction potential here, is truncated at $r=10\sigma$. The NPs also experience a torque force, which for completeness is reported in Appendix~\ref{app::torque}. 

The nature of the third component determines $\phi_{33}^{\rm tot}$, similar to the cross interaction potentials $\phi_{13}$ and $\phi_{23}$ between the third component and the polymer species P$_1$ and P$_2$, respectively. For a general overview of the effective interactions in soft matter systems see, e.g., Refs.~\citenum{belloni2000colloidal,likos2001effective,likos2006soft,narros2010influence,mcmanus2016physics}. Here, we have colloidal NPs dispersed in a two polymer component aqueous solution.  Notably, species affinities are generally asymmetric, i.e. in three solute component dispersions, the affinity of any solute species favors interactions with one of the other two components. Also in Ref.~\citenum{rigoni2022ferrofluidic}, the NPs mostly partition in the dextran rich phase due to favorable interaction between dextran and the citrate groups coating of the iron oxide NPs. 

Let us next consider possible cross interaction potentials $\phi_{13}$ and $\phi_{23}$. Reference~\cite{bolhuis2002derive} reports a parameterised effective potential for colloid-polymer mixture. Nevertheless, in theoretical approaches, such as in Refs.~\cite{dijkstra1999phase,binder2014perspective,stopper2016structural}, the interactions in the context of colloid-polymer mixtures are commonly modelled via the so called Asakura–Oosawa–Vrij potential~\cite{asakura1954interaction}. In it, the interactions between colloids and polymers are modelled via a hard-sphere potential. Here, we use more realistic softer potential forms. To match qualitatively the asymmetricity in affinity response of Ref.~\cite{rigoni2022ferrofluidic}, P$_{1}$ and NPs are set to interact via a purely repulsive WCA-type potential. The form of Eq.~(\ref{WCA}), with $\sigma_{13}=\sigma$ is assumed. In practice, this potential models steric interactions, i.e. corresponds to a ``neutral'' affinity.  The functional form of the interaction between P$_{2}$ and the third component is unknown. Generally, the mechanisms behind the partitioning in ATPSs~\cite{rigoni2022ferrofluidic} are complex~\cite{helfrich2005partitioning,long2006nanoparticle,asenjo2011aqueous,iqbal2016aqueous}. However, one may assume the attractive interaction to be short ranged, due to screened electrostatic interactions. For these reasons, we opt for a standard Mie potential~\cite{mie1903kinetischen} of the form
\begin{equation}\label{eq::mie}
\phi^{\rm Mie}_{23}(r)=4\varepsilon_{\rm 23}\left[\left(\frac{\sigma}{r}\right)^{48} - \left( \frac{\sigma}{r}\right)^{24}\right].
\end{equation}
Here, the diameter of P$_{2}$ is set to be $\sigma$, and the geometrical average is used to obtain the cross interaction range. The level of affinity is controlled by $\varepsilon_{23}$. Because the interaction form in real systems is unknown, connecting the value of $\varepsilon_{23}$ with experimental setups is arduous. However, exploration of different values vs partitioning response in actual systems can be used to match the model with experimental systems (see below). Previously, potentials similar to Eq.~(\ref{eq::mie}) have been used to model short-ranged interactions between colloids~\cite{rouwhorst2020nonequilibrium}, and can be potentially used to model short-ranged interactions for globular protein crystallization~\cite{foffi2002phase}.

To examine the NP partitioning response, we consider a system where $5000$ NPs ($\rho_{\rm NP}\sigma^3\approx0.093$) with identical (reduced) dipole moment strength~\cite{comment_on_B_field} $\mu_i=\mu=2\>{\rm for}\>i=1,..., 5000$ are added to A1 (see Table~\ref{table::1}). Assuming $\sigma=10$ nm, the choice of dipole moment value is equivalent to approximately $4\times 10^{-19} {\rm Am}^2$. For comparison, {\it Sample 0} in Ref.~\citenum{rigoni2022ferrofluidic}, assuming same $\sigma$ value, contains approximately $2250$ NPs in the same volume (see Appendix~\ref{app::numbers}). This is equivalent to $\rho_{\rm NP}^{\rm exp}\sigma^3\approx0.042$, with dipole moments of approximately $7\times 10^{-20} {\rm Am}^2$. For the latter, the average diameter of NPs in the experimental system was estimated to be 7 nm. 

The presence of the NPs in the system changes the conditions for phase separation for two reasons. First, the NPs exclude volume available to the polymers. This alters the thermodynamic state point. Second, the attractive interaction of the NPs with P$_{2}$ via Eq.~(\ref{eq::mie}) locally enhances the aggregation of P$_{2}$. Hence, the choice of the value of $\varepsilon_{23}$ is important for the system response. To map and demonstrate the effect of this choice, we investigate the system response for $\beta\varepsilon_{23}=1,3$, and 4, corresponding to systems B1, B2, and B3, respectively; system composition and parameter details in Table~\ref{table::1}. Figure~\ref{figure::2} shows, by simulation snapshot visualizations and the corresponding radial distribution function data, the configurations adopted for different values of $\varepsilon_{23}$. Snapshots correspond to configurations after $50\times 10^6$ time steps, while the radial distribution function data are obtained by averaging over all particles and over the last $10\%$ of the simulations. In system B1, the NPs remain relatively homogeneously distributed and present in both phases. The radial distribution function corresponding to NP and the affinity polymer species P$_2$, $g_{23}$ shows a small peak, but no long-range correlation. Long-range correlations are also absent in $g_{33}$. On the other hand, system B2 presents a stronger phase separation between the two polymer phases, mediated by the attraction between NPs and P$_{2}$. The NPs partition dominantly into the phase rich in P$_{2}$. This is quantified by the clear long-range correlation in both $g_{23}$ and $g_{33}$. Finally, in system B3, the NPs segregate very strongly, together with P$_{2}$, into one phase. This is evident both from panel (c) and from the corresponding radial distribution functions in panels (d) and (e). This setup has very slow dynamics, including NPs jamming, i.e. the NPs not behaving as a liquid. For modelling ATPS response, the affinity set for system B3 is thus too high, indicating the existence of an upper limit for the value of $\varepsilon_{23}$ to result in liquid-like behavior in the system. Basically, the performed mapping indicates a range over which asymmetric affinity in this type of three solute component system can vary to result in modestly preferential to highly selective partitioning, while still pertaining fluid character.

\subsubsection*{Polymer-NP mixture under gravity-like conditions}\label{sec::grav}

\begin{figure}[h!]
    \centering
    \includegraphics[width=\linewidth]{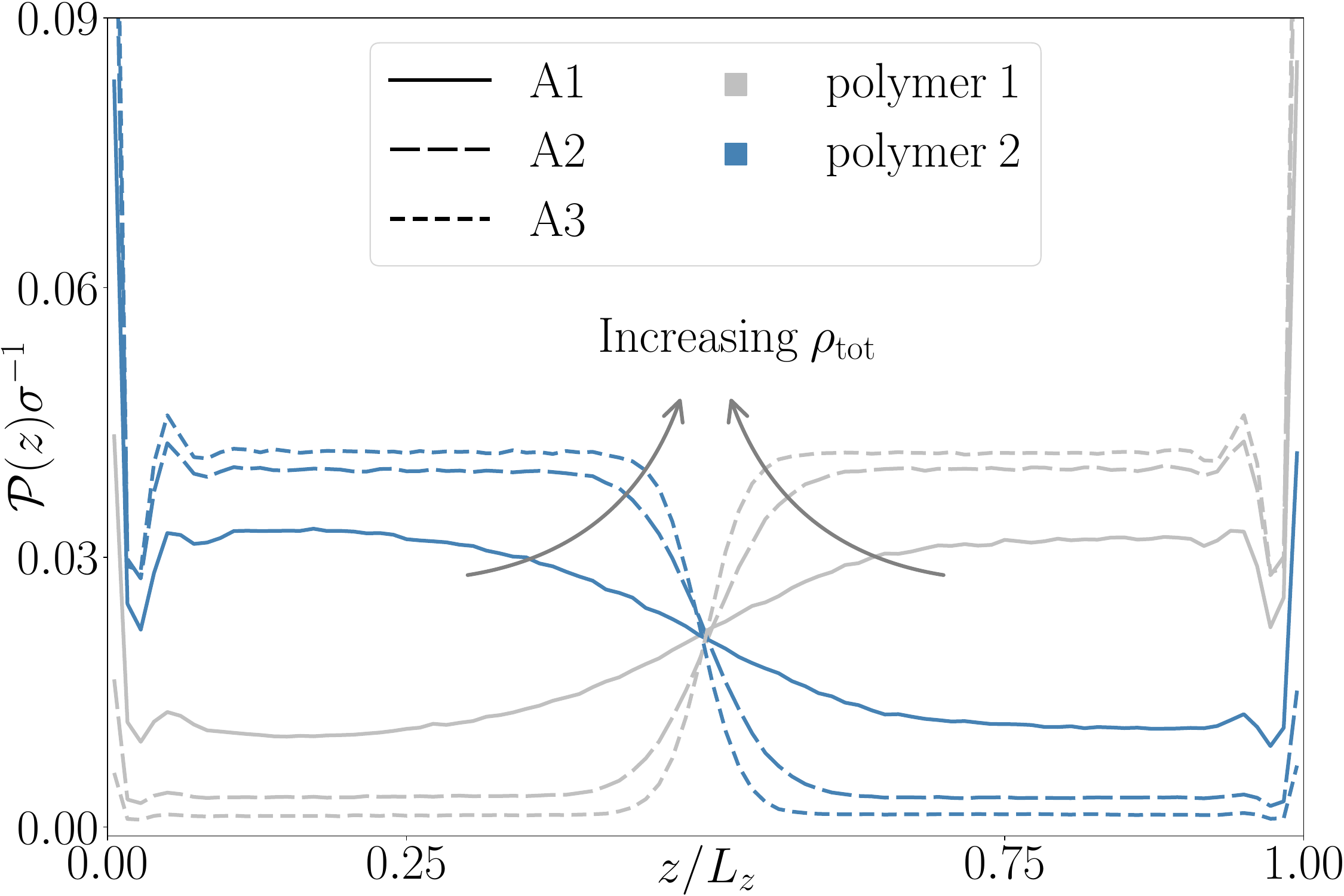}
    \caption{Probability distribution of finding  P$_1$ and P$_2$ in systems A1, A2, and A3. Gravity effects are considered along the $z$ axis, and the data is averaged over the production runs. Sketched arrows show the data trend with increasing total particle number density in the system.}
    \label{figure::3}
\end{figure}

The data of the previous section indicates that the phase separation is driven by polymer incompatibility, and influenced by additional species via the affinities. To have a modelling description in closer match with real systems, we next consider gravitational effects. Motivated by the reference system~\cite{rigoni2022ferrofluidic}, P$_{2}$ and the NPs are set to be heavier than P$_{1}$. To model this, we initialize the systems by setting all P$_{2}$ and NPs in the bottom half of the system, while all P$_{1}$ occupy the region corresponding to the top half. Gravity is set to act along the $z$ axis. Note that no explicit gravitational force is added to the model, but rather the system is allowed to evolve from this configuration. This is justified by the fact that the capillary length of the two-phase system in Ref.~\citenum{rigoni2022ferrofluidic} is on the order of tens of $\mu$m, whereas the linear dimension of the simulated system is on the order of a $\mu$m. The systems are equilibrated for $150\times 10^6$ time steps when NPs are present, and for $300\times 10^6$ time steps when the solution contains only polymers. After this, production runs of respectively $30\times 10^6$ and $150\times 10^6$ time steps are carried out. 
Figure~\ref{figure::3} presents the resulting $z$-axial probability distributions $\mathcal{P}$ for finding the different components in systems A1, A2, and A3, corresponding to the polymer - polymer systems summarized in Table~\ref{table::1}. The data is calculated by averaging over the production runs. $\mathcal{P}$ is normalized following
\begin{equation}
\int_0^{L_z}\mathcal{P}(z)dz=1.
\end{equation}
The data shows that the degree of polymer P$_1$ and P$_2$ mixing decreases as the concentration of the polymers $\rho_{\rm tot}$ increases, as expected. Nevertheless, all examined systems exhibit a clear interface between the two phases at $z= L_z/2$. For system A1, the interface width is quite broad as the system is close to the critical concentration for phase separation. Unsurprisingly, $\mathcal{P}(z)$ is perfectly symmetric in terms of P$_1$ and P$_2$. The accumulation peaks at the walls are due to packing effects at the reflective (hard) wall. 
\begin{figure}[h!]
    \centering
    \includegraphics[width=\linewidth]{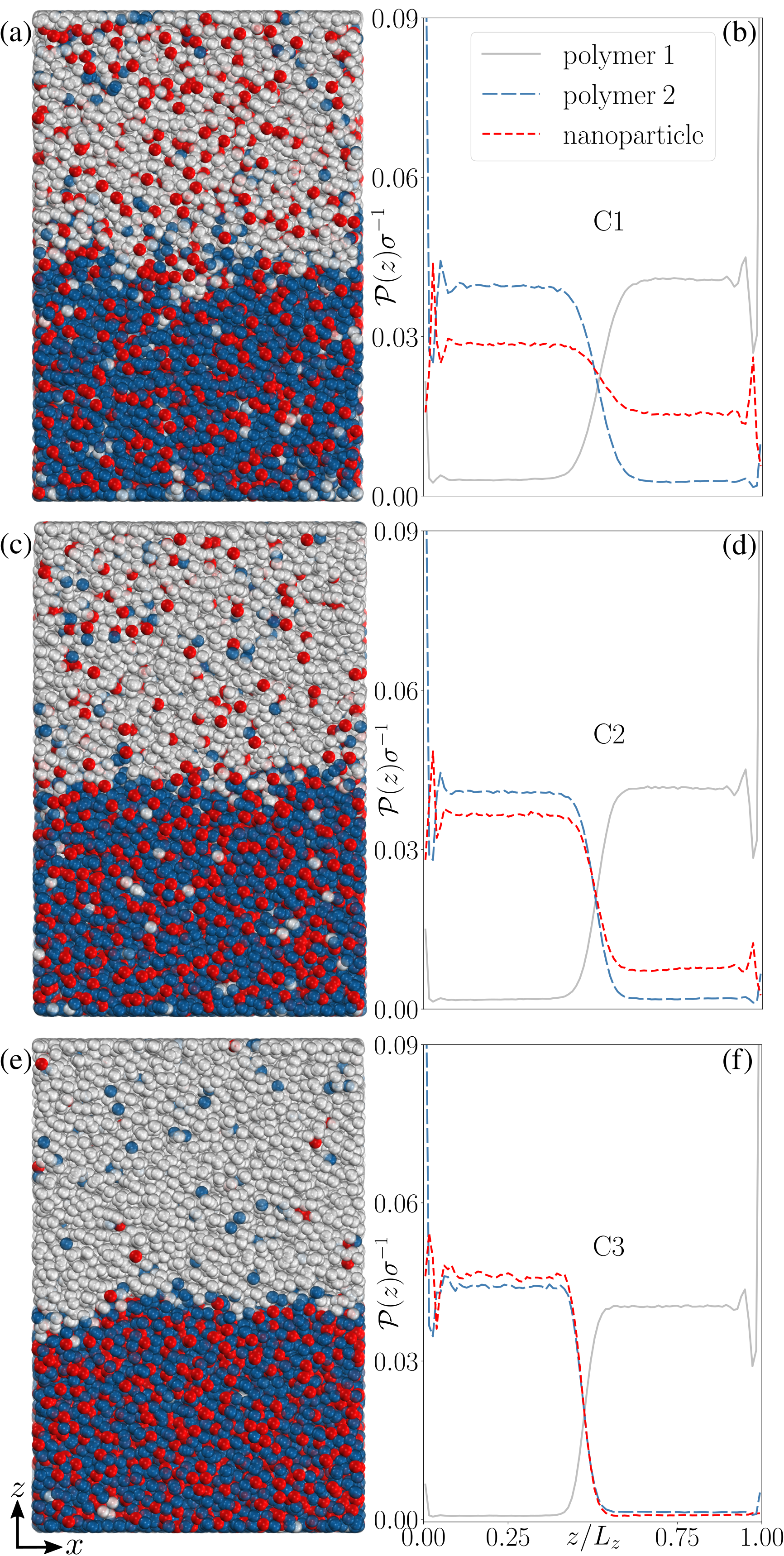}
    \caption{Assembly configurations cross section visualizations (left column) and corresponding $z$-axial probability distributions of the different components (right column) for C1, C2, and C3 (top to bottom). P$_1$ are grey, P$_2$ blue, and NPs red. }
    \label{figure::4}
\end{figure}
For examining the effect of gravity on polymer - polymer - nanoparticle systems, we choose the case with an intermediate total polymer concentration, system A2, and consider the effect of 5000 NPs. As before, the affinity towards P$_2$ is varied. Values $\beta\varepsilon_{23}=1,2$ and 3 are referred to as systems C1, C2, and C3, see Table~\ref{table::1}. Figure~\ref{figure::4} shows the cross section visualizations corresponding to the configurations at the end of the production runs, and the corresponding probability distributions $\mathcal{P}(z)$ over its entirety. C1, in which the NPs have low affinity to P$_2$, panels (a) and (b), exhibits a practically identical polymer distribution as in the corresponding system without NPs (system A2, Fig.~\ref{figure::3}). The NPs are distributed slightly asymmetrically, due to small interaction preference with P$_2$. 

Here, the $z$-axial distribution can be used to determine the phase boundary between the P$_1$ and P$_2$ rich phases. We consider the phase boundary to be the point where $\mathcal{P}(z)$ of P$_1$ equals the one of P$_2$. Let us call the ratio of NPs in the P$_2$ phase $\varphi$. For system C1, we find $\varphi \approx 63 \%$, leaving the remaining $37 \%$ of NPs in the P$_1$ phase. Increasing affinity, in system C2 (see panels (c) and (d)), leads to a more pronounced partitioning of the NPs, and a value of $\varphi\approx 81 \%$. The system with highest affinity studied here, C3, (see panels (e) and (f)), results in an even stronger asymmetry of the distribution of the NPs with $\varphi\approx 96 \%$, and only $4 \%$ of the NPs in the P$_1$ rich phase. Additionally, the interface moves towards the P$_2$ rich phase, resulting in the coexisting density of P$_2$ increasing, i.e. both P$_2$ and NPs concentrations increase in the P$_2$ phase. In other words, system C3 demonstrates a case where the NPs enhance phase separation of the polymers, but also affect the packing density in their affinity phase. For completeness, we also calculate the value of $\log_{10}P$, as is commonly measured in experiments. The values are shown in Table~\ref{table::2}.

Interestingly, the partitioning of the NPs in Ref.~\citenum{rigoni2022ferrofluidic}, measured by estimating the NP concentration in each phase by probing the local density, light absorption, and the magnetic properties, varies, depending on the level of dilution of the samples, between $91 \%$ and $94 \%$ of the NPs in their affinity phase, i.e. the dextran rich phase. The small changes in partitioning, in a system close to the phase separation critical point, are a signal of a strong preference of the NPs to be close to the dextran molecules in the examination setup of Ref.~\cite{rigoni2022ferrofluidic}. 

Our findings allow concluding that at least two handles for controlling the partitioning of a third component in an ATPS exist. First, the thermodynamic state point, which controls the coexisting densities of the polymer mixture provides a direct handle. The second is controlling the coupling between the density distribution of the polymers and the partitioned component, which in the model takes place via $\beta\varepsilon_{23}$, and in the experiments through the affinity between the partitioning component and the two polymers. The constructed simple model has direct access to both means of control, and has thus the potential to provide prediction trends for partitioning in an easily verifiable way as demonstrated by the experimental results, see next section. 

An important connection point with actual ATPSs, is the (reduced) surface tension~\cite{atefi2015interfacial} between the P$_1$ and P$_2$ rich phases:
\begin{equation}
    \gamma^*=-\int_{z_1}^{z_2}\left[ \langle\varsigma_{zz}(z)\rangle-\frac{\langle\varsigma_{xx}(z)\rangle+\langle\varsigma_{yy}(z)\rangle}{2}\right]dz,
\end{equation}
where $\varsigma_{ii}$ for $i=x,y,z$ are the diagonal components of the stress tensor. For calculating $\gamma^*$, we set $z_1=L_z/4$ and $z_2=3L_z/4$ (except for the system A1, where $z_1=L_z/6$ and $z_2=5L_z/6$ due to the broad interface). A comparison of the resulting surface tension estimates for all considered systems is summarized in Table~\ref{table::2}. In this, the surface tension $\gamma=\gamma^* k_{\rm B}T\sigma^{-2}$ is calculated for $T=298$ K and $\sigma=10$ nm.
\begin{table}[h!]
\centering
\begin{tabular}{ C{1.2cm} C{1.3cm} C{1.4cm} C{1.6cm} C{2cm}}
System & $\gamma^*$ & $\gamma\>(\mu$N/m) & $\varphi$ & $\log_{10}P$\\  \hline\hline
A1 & $0.004$ & $0.16$ & - & -\\ \hline
A2 & $0.051$ & $2.10$ & - & -\\ \hline
A3 & $0.125$ & $5.14$ & - & - \\\hline
C1 & $0.051$ & $2.10$ & $63\pm 1\%$ & $0.27 \pm 0.01$\\ \hline
C2 & $0.080$ & $3.29$ & $81\pm 1\%$ & $0.68 \pm 0.01$\\ \hline
C3 & $0.115$ & $4.73$ & $96\pm 1\%$ & $1.78 \pm 0.02$\\ \hline
E1 & - & - & - & $0.81\pm 0.01$\\ \hline
E2 & - & - & - & $0.85\pm0.01$\\ \hline
E3 & - & - & - & $1.05\pm 0.01$\\ 
\end{tabular}
\caption{Dimensionless surface tensions $\gamma^*$ and example values $\gamma$ for $T=298$ K and $\sigma=10$ nm. $\varphi$ is the ratio of NPs partitioned in the P$_2$ phase. $\log_{10}P$ has been calculated in the simulation systems based on the ratio of the plateau values of the NPs densities in the two phases and from Eq.~(\ref{eq::logP}) in the experiments of the current work. In Ref.~\citenum{rigoni2022ferrofluidic}, values of $\gamma$ vary between 0.45 and 0.9 $\mu$N/m, and the ratio between the saturation magnetizations led to $\log_{10}P$ between $1.22\pm0.09$ and $1.07\pm0.08$, depending on the dilution of the sample.} \label{table::2}
\end{table}
The data shows that the addition of NPs with relatively low affinity to P$_2$ ($\beta\varepsilon_{23}=1$) does not have a significant effect on the surface tension. On the other hand, increasing the affinity leads to the surface tension increasing by up to a factor of $\approx 2.3$ for $\beta\varepsilon_{23}=3$. These values are well in the range of surface tensions commonly measured experimentally for ATPSs~\cite{atefi2014ultralow, rigoni2022ferrofluidic}. The surface tension is also expected to depend on the NP concentration, variation not explored in this work. Also, since $\gamma=\gamma^* k_{\rm B}T\sigma^{-2}$ and the value of $\sigma$ in real units is not fixed in the modelling, it is evident that the surface tension can be controlled by changing the size of the polymers. This relation has been found also in colloid-polymer binary mixture, yet as a function of colloidal diameter~\cite{brader2000fluid}. Additionally, the surface tension also depends on the difference between the coexisting densities in the two phases, vanishing at the critical point~\cite{archer2001binary}.

Even though the presented polymer - polymer - nanoparticle modelling approach was developed for capturing partitioning in ATPSs at general level, the range of values obtained for the surface tension using a $\sigma=10$ nm, a realistic value based on Ref.~\cite{rigoni2022ferrofluidic}, are rather satisfactory. However, the simplification choice $R_{ii}=R$, for $i=1,2$ does not correspond to the size ratio of dextran and PEG in Ref.~\citenum{rigoni2022ferrofluidic}. Notably, the phase diagram changes based on different choices of $R_{ij}$~\cite{archer2001binary}; This means that the phase diagram variation cannot be generalized as a function of $R_{ij}$. Furthermore, for more precise predictions of both partitioning and surface tension, additional effort should be put into fitting the thermodynamic state point of the model to fit the coexisting phase density of the experimental setup. 

\subsection*{ATPS partitioning results in experiments}\label{sec::exp}
\begin{figure*}[!htb]
    \centering
    \includegraphics[width=\linewidth]{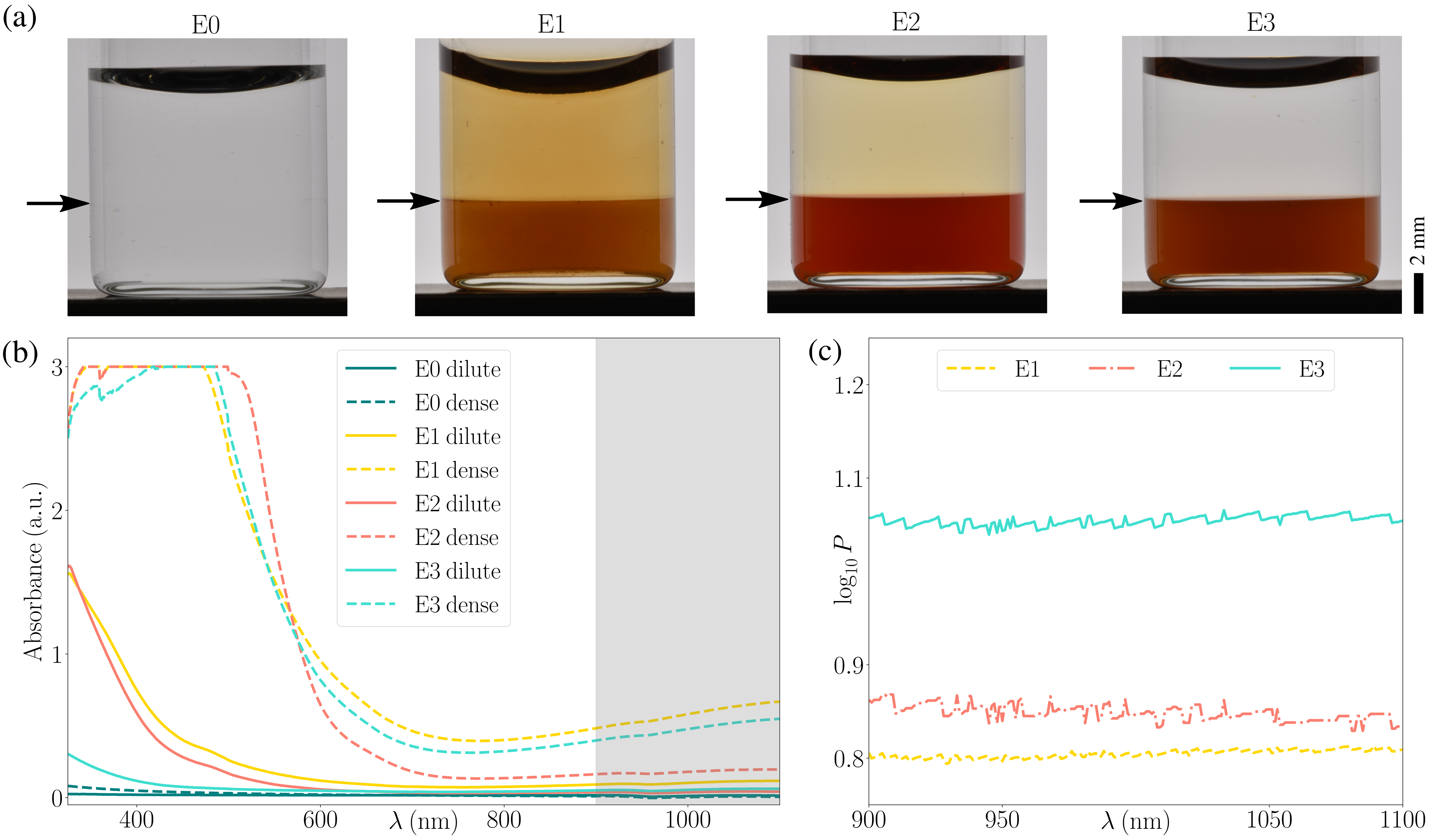}
    \caption{(a) Photographs of the vials used for the spectrophotometry experiments for systems E0, E1, E2, and E3. Arrows indicate the position of the interface in each system. (b) Raw data profiles of the light absorption as a function of the wavelength $\lambda$ for both dense and dilute phases of each sample. (c) Calculated $\log_{10} P$ value for the dispersions of NPs in ATPSs E1, E2, and E3 as a function of $\lambda$ in the range corresponding to the shaded region in panel (b).}
    \label{figure::exp}
\end{figure*}
Analogous variation of NP affinity to the polymers as in the simulation results of Fig.~\ref{figure::4} was obtained in an experimental setup by varying the coating of iron oxide NPs in a PEG-dextran ATPS. The NP-free PEG-dextran ATPS is referred to as system E0. We selected the following NPs: PBG300 pegylated iron oxide NPs from Ferrotec (E1), synthesised dispersion of citrated coated iron oxide NPs~\cite{rigoni2022ferrofluidic} (E2), and 45-111-701 phospholipid coated iron oxide NPs from Micromod (E3). Due to the affinities of the NPs with PEG and dextran differing, the partitioning in systems E1, E2, and E3 is visibly different (see Fig.~\ref{figure::exp}(a)). A quantitative estimation of the partitioning was obtained by the light absorption of the dilute and dense phases (in terms of NPs concentration) of each sample, measured using a spectrophotometer (see Fig.~\ref{figure::exp}(b)). After subtracting the signal contribution rising from the polymers, the partitioning coefficient $P$, commonly used to characterise partitioning within ATPSs~\cite{iqbal2016aqueous}, was evaluated using the Beer-Lambert law for light absorption:  
\begin{equation}\label{eq::logP}
    \log_{10} P = \log_{10} \left(A_{\rm dense}/A_{\rm dilute}\right),
\end{equation}
where $A_{\rm dense}$ and $A_{\rm dilute}$ are the (wavelength dependent) absorption of the dense and the dilute phase, respectively (see Ref.~\citenum{rigoni2022ferrofluidic}, Supplementary note 6)). The logarithmic form connects directly to the energetics of the system, such as chemical potential, enthalpy, entropy, and Gibbs free energy~\cite{grilo2016partitioning}. The calculated values of $\log_{10} P$ at different wavelengths $\lambda$ (Fig.~\ref{figure::exp}(c)) were averaged to obtain the partitioning as $0.81\pm 0.01$ for E1, $0.85\pm0.01$ for E2, and $1.05\pm 0.01$ for E3 (see Table~\ref{table::2}).

\subsection*{Effect of external magnetic field on NP-ATPS by computational model}\label{sec::mag}
\begin{figure}[h!]
    \centering
    \includegraphics[width=0.84\linewidth]{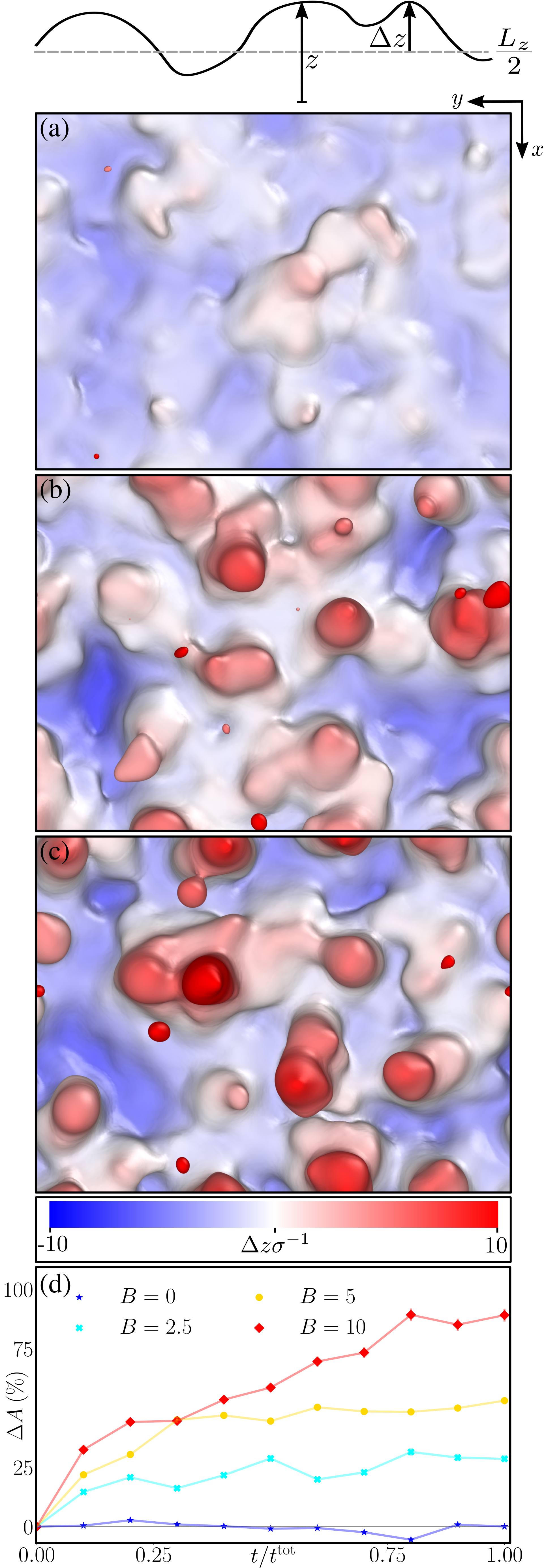}
    \caption{Height changes of the bottom phase for C3 (a) before turning on $B$, (b) after $50\times 10^6$ time steps with $B=5$, and (c) with $B=10$. In (d), relative surface area change $\Delta A$ over time, for different values of $B$.}
    \label{figure::5}
\end{figure}
Let us next consider the polymer mixture and NPs model under gravity-like conditions in external magnetic field. The equilibrated system C3, Fig.~\ref{figure::4}(e), is considered as it best matches with the experimental reference setup~\cite{rigoni2022ferrofluidic}. An external magnetic field $\textbf{B}=B\textbf{e}_z$, where $B$ is the reduced magnetic field strength~\cite{comment_on_B_field} is applied on the system. In addition to the dipole-dipole interactions (force and torque) introduced in Sec.~\ref{sec::mix_and_np}, a further torque force on the magnetic NPs has to be considered. This torque has the form $\textbf{T}^{\rm B}=\boldsymbol{\mu}\times \textbf{B}$. Since $\textbf{B}$ is uniform in the $z$-direction, $\textbf{T}^{\rm B}$ is the only force on the NPs rising from introducing the external magnetic field. The charges of the NPs are considered via Eq.~(\ref{eq::electro}) as effective contributions. This is naturally an approximation, but allows excluding the magnetic field-charge contributions. If the magnetic field is strong enough, the dipole moments align in the direction of $\textbf{B}$ (here the $z$ axis). This gives rise to long range dipole-dipole correlations. One can expect a modulation of the interface, potentially resulting in a variety of patterns. 

For a dilute ferrofluid and small magnetic field, the relative permeability of the bottom phase is $\mu_r = 1+\mu_0\rho_{\rm B}\mu^2(k_{\rm B}T)^{-1}$. In this, $\mu_0$ is the permeability of vacuum and $\rho_{\rm B}$ is the number density of NPs in the ferrofluid phase~\cite{rosensweig2013ferrohydrodynamics}. In system C3, $\rho_{\rm B}\approx 5000\times 0.96\> / \left(30\times 40\times 22.5\>\sigma^3\right)\approx 0.18\>\sigma^{-3}$. This corresponds to relative permeability $\mu_r \approx 10$. The classical Rosenzweig instability~\cite{rosensweig2013ferrohydrodynamics}, in the absence of gravitational forces, results in interfacial oscillations above a characteristic length given by
\begin{equation}
    \lambda^* = \frac{\gamma \mu_0 \mu_r (1+\mu_r)}{(\mu_r-1)^2B^2}.
\end{equation}
The surface tension $\gamma$ of system C3 and a choice of representative value of the magnetic field strength of $B=1$ (in real units corresponding to approximately 16 mT~\cite{comment_on_B_field}) results in a lower limit $\lambda^*\approx 2\sigma$. In light of such a small $\lambda^*$, continuum theory cannot be expected to work quantitatively in this regime. Hence, particle-based simulations are a more appropriate approach to resolve the interface response under these conditions.
Experimentally~\cite{rigoni2022ferrofluidic}, the periodicity of the patterns rising at the interface of this particular ATPS system is on the order of $100-300$ $\mu$m. To the best of our knowledge, these pattern periodicities are the shortest so far measured for Rosensweig-type instability at the interface between continuous phases. The current modelling approach allows, however, to reach scales up approximately $50-100\sigma$ (few $\mu$m, depending on the size of the solutes). Nevertheless, it is worth to address the emergence of pattern at this smaller scale, where the length scale of the fluctuations is on the order of a few particle diameters.

Figure~\ref{figure::5} shows the system response to external magnetic field via the change of height of the interface with respect to $L_z/2$, $\Delta z$, see sketch at the top. The iso-surface map parameters used for this representation are radius scaling 2$\times$, iso-value 3, and both NPs and P$_2$ radii equal to $\sigma/2$. The data shows that before the field is turned on, panel (a), the interface remains relatively flat with oscillations that are on the order of one or two particle diameters. These fluctuations are due to the liquid nature of the system (capillary waves). We then set $B=2.5,5$ and 10, and run simulations for $50\times 10^6$ time steps, corresponding to $t^{\rm tot}$ in Fig.~\ref{figure::5}(d). Figure~\ref{figure::5}(b) shows the interface modulation for $B=5$, with clearly enhanced fluctuations of the interface. A stronger magnetic field, $B=10$, increases these oscillations, as shown in panel (c). Panel (d) quantifies the effect of the magnetic field on the interface by measuring the relative surface area change $\Delta A$ over time. The change is measured with respect to the initial area. The data shows that this change grows over time, reaching an approximately constant value. The saturation value monotonically depends on the strength of $B$. The corresponding change of surface area for $B=0$ is shown for comparison. Here, the fluctuation of the interface in the absence of an external magnetic field is less than $\pm 5\%$ . 

The interface modulation found here is more dynamic than the one observed experimentally~\cite{rigoni2022ferrofluidic}. Additionally, it does not exhibit a periodic structure, opposed to the stationary patterns found experimentally~\cite{rigoni2022ferrofluidic}. Nevertheless, the modelling response shows that this type of coarse-grained model has the potential to extract magnetic self-organization response and interface instabilities. However, the modelled systems sizes remain significantly smaller than the imaging resolution in  experiments.

\section{Conclusions}\label{sec::summary}

In this work, we presented a general coarse-grained Brownian dynamics-based simulation approach to study phase separation of aqueous polymer mixtures and partitioning of an arbitrary third component therein. Motivated by the findings of Ref.~\citenum{rigoni2022ferrofluidic}, the model was employed to investigate the segregation of magnetic NPs and their response to an external magnetic field. The key feature of the approach relies in the effective interaction between the immiscible polymers, which were described by a so-called Gaussian core model, which is particularly accurate for interactions of linear polymers in good solvents~\cite{louis2000can} and also exhibits phase separation behavior~\cite{louis2000mean}. 
Our findings demonstrate that at least two handles for controlling the partitioning of a third component in an ATPS exist. The first is accessible via the thermodynamic state point, which controls the coexisting densities of the polymer mixture. The second is controlled by the affinity asymmetry between the partitioned component and the polymer species. 
Given that partitioning within ATPSs is one of the most important features of their applications~\cite{walter1986partitioning, yau2015current, iqbal2016aqueous}, although the model was not explicitly parametrized to match the reference system~\cite{rigoni2022ferrofluidic}, predictions of partitioning behavior were well within the range of experimentally measured values.
Important for designing ATPSs with desired partitioning response, the modelling work revealed the sensitivity of the system response to interactions parameters (determining the component affinities) and via that, control of the partitioning by them. Such control was achieved experimentally, by varying the surface coating of the NPs to tune the affinity to polymer species. 

Thermodynamic quantities, such as the interfacial tension, can also be obtained within our modelling framework. Small interfacial and surface tensions, present in many biomolecular systems, and more generally in ATPSs, are often challenging to capture experimentally. Overall, the values of the surface tension estimated from the model for the different scenarios of ATPS are well within experimentally measured ranges~\cite{atefi2014ultralow, rigoni2022ferrofluidic}, see also Table~\ref{table::2}. The comparison of the different ATPSs in the computational modelling also revealed the influence of the attraction parameter between the third component and the polymer species on the surface tension of the polymer-polymer interface.

Finally, our particle-based approach is particularly well-suited for studying the response of magnetic particles under external magnetic field in ATPSs. As previously reported for the reference system~\cite{rigoni2022ferrofluidic}, we also observed interface modulations in our simulations. On the other hand, the interface fluctuations observed in the simulations occurred on much smaller length scales, exhibited less periodic structures, and were more dynamic than those in experiments. Nevertheless, the findings demonstrate that the coarse-grained simulation approach has the potential to resolve coupled magnetic self-organization and interfacial instabilities in ATPSs. In closing, we wish to emphasize that our approach to modelling ATPSs with or without a third solute species is readily generalizable to examining the influence of solute species affinities and system compositions on partitioning and the phase separation response of general two-phase solutions. As such, we anticipate several exciting applications of it in future works.   

\section*{Acknowledgments}
This work was supported by the Swiss National Science Foundation under the project no. P500PT$\_$206916 (A.S.) and the Academy of Finland through its Centres of Excellence Programs (2022-2029, LIBER) under projects no. 346111 (M.S.) and 346112 (J.T.). MPH was supported by the National Science Foundation through the Princeton University (PCCM) Materials Research Science and Engineering Center DMR-2011750. A.S. warmly thanks Bob Evans for extensive scientific discussions and for his hospitality during the research visit in Bristol. Computational resources by CSC IT Centre for Finland, the Aalto Science-IT project, and RAMI -- RawMatters Finland Infrastructure are also gratefully acknowledged.

\newpage
\appendix
\section{Validity of the viscous limit in simulations}\label{app::viscous_limit}
Brownian dynamics simulations modelling approach used in this work assumes no average acceleration of the particles. The validity of this approximation depends on the decay time of the velocity time correlation vs the time scale of interest. The velocity time correlation function in Langevin dynamics decays as $\sim \exp[-t/\tau']$. The viscous limit, which implies no velocity correlation, is valid if the time scale of interest, $t$, is much larger than $\tau'=m/\Gamma_{\rm t}$, where $m$ is the mass of the particles and $\Gamma_{\rm t}$ the translational friction coefficient. It is standard to impose that this is valid at time scales corresponding to the self-diffusion time $\tau=\sigma^2/D_{\rm t}=(2R)^2\Gamma_{\rm t}k_{\rm B}^{-1} T^{-1}$. Here, $\sigma$ represents the diameter of the particle while $R$ its radius, $D_{\rm t}$ the translational diffusion coefficient, $k_{\rm B}$ the Boltzmann constant, and $T$ the temperature.

In a viscous fluid, the friction coefficient can be approximated by $\Gamma_{\rm t}=6\pi\eta R$, where $\eta$ is the viscosity of the fluid. In Ref.~\citenum{rigoni2022ferrofluidic}, the NPs are iron oxide, with $m\approx 10^{-21}$ kg, $R \approx 3.5$ nm, and the solvent is water ($\eta_{\rm w}=10^{-3}$ kg m$^{-1}$ s$^{-1}$). This provides a friction coefficient $\Gamma_{\rm t} \approx 6.6 \times 10^{-11}$ kg s$^{-1}$, and $\tau' \approx 1.5 \times 10^{-11}$ s. A value for the self-diffusion time of the reference system of Ref.~\cite{rigoni2022ferrofluidic} hence is $\tau \approx 8 \times 10^{-7}$ s. This means, $\tau/\tau'\approx 5.3 \times 10^4 \gg 1$, which confirms the validity of the viscous limit assumption for the system of Ref.~\citenum{rigoni2022ferrofluidic}. Similar calculation can be done for the experimental systems E1, E2, and E3, but as only the NP coatings differ, the validity of the viscous limit is clear.

\section{Estimation of the number of polymer molecules in experiments}\label{app::numbers}
We estimate the number of molecules (dextran or PEG) in the systems characterized experimentally in Ref.~\citenum{rigoni2022ferrofluidic}, respectively of particles (maghemite nanoparticles), in a volume $V$ corresponding to the simulation box size (here $30\sigma\times 40\sigma\times 45\sigma$), using the ratio
\begin{equation}\label{eq::app-number}
    N_{i}^{\rm exp}\approx\frac{m_i^{\rm exp}}{m_i}\frac{V}{V^{\rm exp}},
\end{equation}
where $m_i^{\rm exp}$ is the mass of the material $i$ in the sample, $m_i$ the mass of a single unit component in that mass, i.e. a single molecule of dextran or PEG, or a single NP, and $V^{\rm exp}$ is the total volume of the sample (40 ml). In Ref.~\citenum{rigoni2022ferrofluidic}, the mass of a single NP is accessible via the density of maghemite (4860 kg m$^{-3}$) and the NP size (volume). Assuming the NPs to be spherical and with a radius of $3.5$~nm, the mass of a single NP is $m_{\rm NP}\approx 87.3\times 10^{-23}{\rm\>kg}$.
On the other hand, the values of $m_i$ for PEG and dextran, as well as all values of $m_i^{\rm exp}$, are taken from the Supplementary Information of Ref.~\citenum{rigoni2022ferrofluidic}. Specifically, $m^{\rm exp}_{\rm NPs}=1.457\times 10^{-3}$ kg, $m^{\rm exp}_{\rm dextran}=1.615\times 10^{-3}$ kg, $m^{\rm exp}_{\rm PEG}=0.740\times 10^{-3}$ kg, $m_{\rm dextran}=83.0\times 10^{-23}$ kg and $m_{\rm PEG}=5.8\times 10^{-23}$ kg. Assuming a value of $\sigma=10$ nm for calculating the volume $V$ of the simulation setup, we get $V=54\times 10^{-21}{\rm\>m^3}$. From Eq.~(\ref{eq::app-number}), the corresponding number of NPs is thus $N_{\rm NPs}^{\rm exp}\approx 2250$, the number of dextran molecules is $N_{\rm dextran}^{\rm exp}\approx 2630$ and finally, the number of PEG molecules is $N^{\rm exp}_{\rm PEG}\approx 17220$.

\section{Dipole-dipole torque in simulations}\label{app::torque}
The torque exerted between the magnetic dipoles of particles $i$ and $j$ are calculated via
\begin{widetext}
\begin{equation}
\begin{split}
\textbf{T}_{ij}(\textbf{r},\boldsymbol{\mu}_i, \boldsymbol{\mu}_j)  &=  -\frac{1}{r^3}\left[1-4\left(\frac{r}{r_{\rm c}}\right)^{\!3} +
3\left(\frac{r}{r_{\rm c}}\right)^{\!4}\right] (\boldsymbol{\mu}_i \times \boldsymbol{\mu}_j) +
 \frac{3}{r^5}\left[1-4\left(\frac{r}{r_{\rm c}}\right)^{\!3} +
3\left(\frac{r}{r_{\rm c}}\right)^{\!4}\right] (\boldsymbol{\mu}_j\cdot\textbf{r})
(\boldsymbol{\mu}_i\times\textbf{r}),\\
\textbf{T}_{ji}(\textbf{r},\boldsymbol{\mu}_i, \boldsymbol{\mu}_j) &=  -\frac{1}{r^3}\left[1-4\left(\frac{r}{r_{\rm c}}\right)^{\!3} +
3\left(\frac{r}{r_{\rm c}}\right)^{\!4}\right](\boldsymbol{\mu}_j \times \boldsymbol{\mu}_i) + \frac{3}{r^5}\left[1-4\left(\frac{r}{r_{\rm c}}\right)^{\!3} +
3\left(\frac{r}{r_c}\right)^{\!4}\right] (\boldsymbol{\mu}_i\cdot\textbf{r})
(\boldsymbol{\mu}_j\times\textbf{r}),
\end{split}
\end{equation}
\end{widetext}
where $\boldsymbol{\mu}$ is the magnetic moment, $\textbf{r}$ the distance vector, $r=\vert\textbf{r}\vert$, and $r_{\rm c}$ the cut off distance. This slightly modified form of the common dipole-dipole interaction torque~\cite{tildesley1987computer} makes the force zero at the cutoff distance, preventing cut off-related problems in energy conservation when integrating the equations of motion over time.

\bibliography{references}

\end{document}